\documentclass[a4paper]{article}
\usepackage{ISCSLP2026}
\usepackage{ifthen}

\usepackage{graphicx}
\usepackage{booktabs}
\usepackage{multirow}
\usepackage{makecell}
\usepackage{array}
\usepackage{adjustbox}
\usepackage{tabularx}
\usepackage{comment}
\usepackage{amsmath}
\usepackage{caption}
\usepackage{float}
\usepackage{url}
\usepackage[skins,breakable]{tcolorbox}
\tcbset{
  enhanced,
  breakable,
  colback=white,
  colframe=black!15,
  boxsep=1ex,
  left=0.6em,right=0.6em,top=0.4em,bottom=0.4em
}

\newboolean{blind}
\setboolean{blind}{false} 

\title{Audio-Maestro: Enhancing Large Audio-Language Models with Tool-Augmented Reasoning}

\name{
  \ifthenelse{\boolean{blind}}{Anonymous to ISCSLP}
  {Kuan-Yi Lee$^1$, Tsung-En Lin$^{1,2}$, Hung-Yi Lee$^1$}
}

\address{
  \ifthenelse{\boolean{blind}}{Anonymous to ISCSLP}
  {
    $^1$ National Taiwan University, Taipei, Taiwan\\
    $^2$ ASUS Open Cloud Infrastructure Software Center, Taipei, Taiwan
  }
}

\email{
  \ifthenelse{\boolean{blind}}{Anonymous to ISCSLP}
  {\{b10901091, b11901154, hungyilee\}@ntu.edu.tw}
}

\begin{document}
\maketitle

% ---------- Abstract ----------
\begin{abstract}
Recent advancements in large multimodal models (LMMs) have shown strong capabilities in audio reasoning. However, most systems rely solely on end-to-end reasoning, limiting interpretability and accuracy for tasks that require structured knowledge or specialized signal analysis. In this work, we present \textbf{Audio-Maestro} -- a tool-augmented audio reasoning framework that enables audio-language models to autonomously call external tools and integrate their timestamped outputs into the reasoning process. This design allows the model to analyze, transform, and interpret audio signals through specialized tools rather than relying solely on end-to-end inference. Experiments show that Audio-Maestro consistently improves general audio reasoning performance: \textbf{Gemini-2.5-flash}'s average accuracy on MMAU-Test rises from 67.4\% to 71.1\%. To our knowledge, Audio-Maestro is the first framework to integrate structured tool output into the large audio language model reasoning process.
\end{abstract}

% ---------- Introduction ----------
\section{Introduction}
Multimodal audio reasoning requires both low-level acoustic analysis and high-level semantic understanding. While end-to-end large multimodal models (LMMs) such as Gemini \cite{team2023gemini} demonstrate remarkable generative ability, they often struggle with domain-specific computations that require structured reasoning—such as chord estimation. These tasks demand not only perception but also symbolic precision.

\begin{figure*}[t]
\centering
\includegraphics[width=1.0\textwidth]{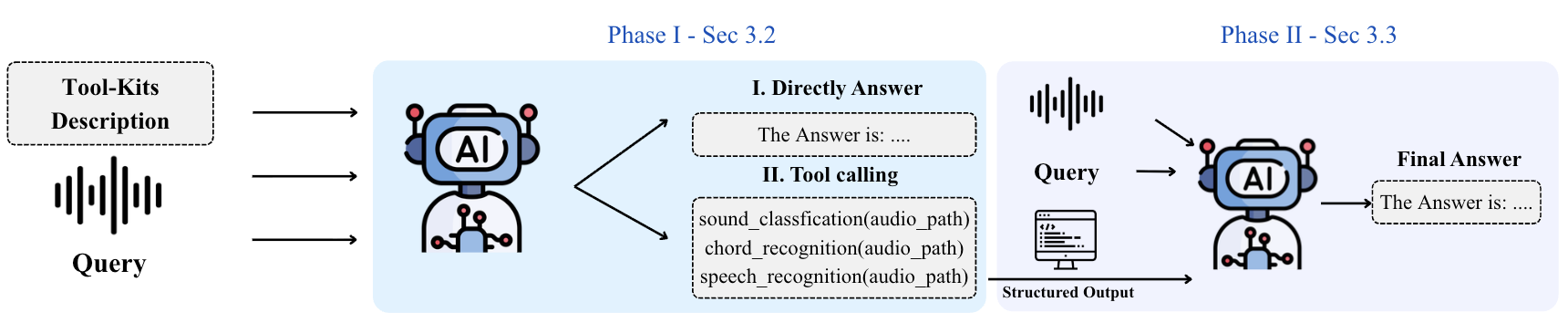}
\caption{\textbf{Overview of the Audio-Maestro framework.} Given an audio input, query, and toolkit, the LALM first decides whether to answer directly or call tools in Phase 1. 
In Phase 2, selected tools are executed on the audio, producing structured, timestamped outputs that are integrated with the query and audio for final inference.}
\label{fig:framework}
\end{figure*}

To address this gap, we introduce \textbf{Audio-Maestro}\footnote{The complete codebase is available at \url{https://anonymous.4open.science/r/Audio-Maestro-FDD6/README.md}.}, as shown in Figure~\ref{fig:framework}, a tool-augmented reasoning framework where a Large Audio-Language Model (LALM) autonomously decomposes complex queries. The LALM decides whether to invoke specialized external tools—such as for chord recognition or speaker diarization—and integrates their structured, time-stamped outputs back into its reasoning process. Our main contribution is extending tool-augmented reasoning to the audio domain, enabling a novel synergy between the LALM's high-level semantic understanding and the tools' precise, low-level acoustic analysis. This approach allows the model to ground its symbolic reasoning in concrete acoustic events, moving beyond monolithic end-to-end representations.

% ---------- Related Work ----------
\section{Related Work}
\subsection{Audio Language Model}
\label{audioreason}
Prior work in audio-language models~\cite{wang2023neural, tangsalmonn, tseng2025taste} provides initial progress toward building general-purpose models. To enhance reasoning capabilities, recent models have incorporated structured inference and reinforcement learning (RL). Mellow~\cite{deshmukh2025mellow} demonstrates lightweight architectures can achieve strong reasoning performance, while Audio-Reasoner~\cite{xie2025audio} employs multiphase pipelines for deeper inference over complex audio scenes. In addition, R1-AQA~\cite{li2025reinforcement} and Omni-R1~\cite{zhong2025omni} further improve reasoning consistency and sample efficiency by fine-tuning LMMs with RL, and Audio-Thinker~\cite{wu2025audio} refines strategies using adaptive rewards.

However, most audio-language models remain end-to-end, relying on implicit internal reasoning. Such systems often require large computational resources and struggle with tasks demanding precise, low-level computations (e.g., pitch extraction). These limitations motivate the exploration of tool-augmented reasoning in section \ref{sec:tool_augmented}.

\subsection{Tool-Augmented Reasoning}
\label{sec:tool_augmented}
Tool-augmented reasoning has emerged as an effective paradigm to extend the capabilities of large language models by delegating specialized subtasks to external modules. Prior work in the language and vision domains—e.g., ART~\cite{paranjape2023art}, Toolformer~\cite{schick2023toolformer}, and visual CoT approaches~\cite{chen2024visualcot}—demonstrates that structured tool calls can improve accuracy and compositional reasoning. 

In the audio domain, some systems have begun to incorporate external components. For instance, Step-Audio 2~\cite{wu2025step} integrates retrieval-augmented generation (RAG) to mitigate hallucination. However, its tool usage is primarily focused on retrieval rather than fine-grained audio analysis. Similarly, systems like Speech-Copilot~\cite{kuan2024speech} and ToolLLM~\cite{qin2024toolllm} have explored tool use, but they primarily interact through text-based interfaces, even when the initial input is speech. However, general audio reasoning tasks often involve precise low-level acoustic computations, which present challenges for existing tool-calling approaches that are primarily designed for symbolic or textual data and may not fully capture the temporal or acoustic structure required for accurate reasoning.

Our framework directly tackles this gap by enabling an LALM to autonomously invoke a diverse toolkit tailored for speech, music, and sound analysis. Unlike prior efforts, we extend the core model to the audio domain, allowing the LALM to leverage precise low-level acoustic details while focusing on high-level semantic understanding

% ---------- Method ----------
\section{Method}

\subsection{Overall Pipeline}
Our framework (Fig.~\ref{fig:framework}) adopts a two-phase design to enable tool-augmented audio reasoning. Given an audio input $x_\text{audio}$ and a textual query $q$, Audio-Maestro aims to combine end-to-end perception with external audio tool execution to generate response. In \textbf{Phase 1}, the model decides whether it can directly answer the question or requires tool assistance. 
In \textbf{Phase 2}, if tools are invoked, the model integrates their structured results back into its reasoning process to form a final response.

This design allows the model to perform context-aware and explainable reasoning over complex audio scenes, bridging low-level perception and high-level symbolic analysis.

\subsection{Phase 1: Decision-Making}
\label{stage1}
Given an input pair $(x_\text{audio}, q)$ and the available tool set $\mathcal{T} = \{t_1, t_2, \dots, t_K\}$, the large audio-language model (LALM) decides whether to produce a direct answer or to invoke one or more external tools. We denote the decision by
\[
a_\text{decision} = \mathcal{M}_\text{LALM}(x_\text{audio}, q, \mathcal{T}) \in \{Ans, C\},
\]
where $Ans$ indicates that the model answers the query directly, and $C$ indicates that one or more tools from $\mathcal{T}$ are called. 

This process follows the paradigm of tool-augmented reasoning \cite{schick2023toolformer,paranjape2023art}, extended to multimodal audio understanding. The model’s decision reflects both semantic understanding and acoustic cues — for instance, detecting emotion shifts, overlapping speakers, or non-speech sounds that might require specialized analysis.

\subsection{Phase 2: Execution and Integration}
\label{stage-2}
If tool calls are triggered, each selected tool $t_k \in \mathcal{T}_\text{sel} \subseteq \mathcal{T}$ is executed on the same audio input:
\[
y_k = t_k(x_\text{audio}).
\]
Each tool produces structured timestamped output, which captures interpretable aspects such as emotion trajectories, sound event durations, or chord progressions. These outputs are serialized and concatenated with the original audio and query representation to form an enriched context:
\[
c_\text{aug} = \text{Concat}(x_\text{audio}, q, y_1, \dots, y_{|\mathcal{T}_\text{sel}|}).
\]
Finally, the LALM generates the answer conditioned on the augmented context:
\[
r = \mathcal{M}_\text{LALM}(c_\text{aug}).
\]
This phase enables the model to incorporate explicit acoustic evidence and symbolic reasoning into response generation, significantly improving interpretability and robustness in complex auditory scenarios.

\subsection{Implementation Details}
\label{sec:implementation_details}
Our framework follow a zero-shot setting, guiding the LALM to autonomously invoke tools via a structured prompt without any task-specific fine-tuning. The prompt consists of three components: a system instruction defining the model's role as an audio expert, detailed descriptions of the available tools, and the user's audio file and text query.

If a tool is invoked, our framework executes it externally and returns the output as a structured JSON string. This structured, timestamped output is then combined with the prompt and the feedback to the LALM to synthesize its final, tool-informed response.

% ---------- Experiment ----------
\section{Experiments Setup}
\subsection{Evaluated Backbones}
To ensure a fair and informative comparison, we select models that exhibit strong reasoning and tool-invocation capabilities. We include \textbf{DeSTA-2.5} \cite{lu2024desta}, \textbf{Gemini-2.5-flash} \cite{team2023gemini}, and \textbf{GPT-4o} \cite{hurst2024gpt}. 

\subsection{Benchmark}
We adopt Massive Multi-Task Audio Understanding and Reasoning (MMAU)\cite{sakshi2024mmau} as our benchmark, it evaluates multimodal audio understanding models on tasks requiring expert-level audio knowledge and complex reasoning, beyond simple classification or transcription. It spans three domains—\textbf{speech, environmental sounds, and music}, and the model need to choose the option from multiple choices. We report the accuracy as evaluation metric.

\subsection{Toolkit}
To support generalizable audio reasoning, we construct a set of domain-specific tools following the Speech-Copilot\cite{kuan2024speech}. Instead of manually designing each function, Speech-Copilot prompt GPT-4o to automatically generate tool interfaces and documentation. Compared with manually curated tool sets, this approach ensures that the tools exhibit low redundancy and high extensibility for various tasks.

In our experiments, we generate 12 tools as Table \ref{tab:toolkit} with Speech-Copilot, which cover key aspects such as diarization or chord recognition. This toolkit serves as the functional backbone of the inference phase described in Section~\ref{stage-2}. Each tool is designed to return structured, timestamped output, enabling the model to align symbolic reasoning with acoustic events in the original audio. 

\begin{table}[t]
\centering
\caption{Tasks decomposed and tools automatically generated with Speech-Copilot~\cite{kuan2024speech} framework, and we select models for each tool.}
\label{tab:toolkit}
\resizebox{\columnwidth}{!}{%
\begin{tabular}{|l|l|}
\hline
\textbf{Modules} & \textbf{Underlying Model/Library for Each Tool} \\
\hline
Speech Recognition & Whisper-large-v3 \cite{whisper} \\
Emotion Recognition & emotion2vec+large \cite{ma2023emotion2vec} \\
Speaker Diarization & pyannote/speaker-diarization-3.1 \cite{pyannote} \\
Speech-to-Noise Ratio  & Brouhaha \cite{lavechin2023brouhaha} \\
Sound Classification & AST \cite{gong2021ast} \\
Sound Duration Analysis & AST (sliding window) \cite{gong2021ast} \\
Melody Recognition & librosa piptrack \cite{McFee2015librosaAA} \\
Chord Recognition & autochord \cite{bayron2021autochord} \\
Chord Duration Analysis & autochord  \cite{bayron2021autochord} \\
Genre Analysis & AST \cite{gong2021ast}, librosa \cite{McFee2015librosaAA} \\
Stress Analysis & MFA \cite{mfa2017}\\
Audio Feature Extraction & librosa \cite{McFee2015librosaAA} \\
\hline
\end{tabular}
}
\end{table}

% ---------- Results ----------
\begin{table*} 
\small 
\caption{Comparison with state-of-the-art audio reasoning models on MMAU. The highest accuracy in each section is in \textbf{bold}. \textbf{Audio-Maestro} consistently outperforms recent reasoning-enhanced models (Audio-Reasoner, R1-AQA) on the MMAU benchmark, demonstrating the efficacy of external tool augmentation.}
\label{tab:main-result}
\centering
\begin{tabular}{l|cccc|cccc}
\toprule
& \multicolumn{4}{c|}{\textbf{Test (Full)}} & \multicolumn{4}{c}{\textbf{Test-mini}} \\
\textbf{Model} & Sound & Music & Speech & Avg & Sound & Music & Speech & Avg \\
\midrule
\multicolumn{9}{c}{\textbf{SOTA Audio Reasoning Models (Internal Reasoning/RL)}} \\
\midrule
Audio-Thinker (Qwen2.5-Omni) \cite{wu2025audio} & \textbf{76.30} & 66.63 & \textbf{73.27} & \textbf{72.83} & 77.48 & 70.36 & 73.37 & 73.70 \\Audio-Reasoner \cite{xie2025audio} & - & - & - & - & 60.06 & 64.30 & 60.70 & 61.71   \\R1-AQA \cite{li2025reinforcement} & 69.76 & 61.40 & 62.70 & 64.36& 68.77 & 64.37 & 63.66 & 65.60  \\\midrule
\multicolumn{9}{c}{\textbf{DeSTA-2.5}} \\ 
\midrule
Text Only + Tool & 57.03 & 52.73 & 48.07 & 52.61 & \textbf{65.17} & 51.20 & 63.96 & 60.11 \\
Audio Without Tool & 63.54 & 55.30 & 61.72 & 60.19 & 63.10 & 54.14 & 68.30 & 61.83 \\
Audio-Maestro & \textbf{63.63} & \textbf{55.34} & \textbf{70.21} & \textbf{63.06} & 64.56 & \textbf{57.48} & \textbf{73.27} & \textbf{65.10} \\
\midrule
\multicolumn{9}{c}{\textbf{Gemini-2.5-flash}} \\
\midrule
Text + Tool & 54.86 & 51.35 & 68.50 & 58.24 & 63.66 & 58.68 & 74.77 & 65.73 \\
Audio Without Tool & 69.50 & 64.40 & 68.27 & 67.39 & 73.27 & 65.57 & 76.58 & 71.86 \\
Audio-Maestro & \textbf{75.19} & \textbf{65.56} & \textbf{72.51} & \textbf{71.09} & \textbf{78.68} & \textbf{69.16} & \textbf{80.48} & \textbf{76.16} \\
\midrule
\multicolumn{9}{c}{\textbf{GPT-4o}} \\
\midrule
Text Only + Tool & 55.46 & 46.68 & 68.27 & 56.80 & 58.49 & 49.86 & 66.52 & 58.29 \\
Audio Without Tool & 63.20 & 49.93 & 69.33 & 60.82 & 64.56 & \textbf{56.29} & 66.67 & 62.50 \\
Audio-Maestro & \textbf{65.98} & \textbf{52.22} & \textbf{73.34} & \textbf{63.85} & \textbf{66.06} & 55.88 & \textbf{72.76} & \textbf{64.83} \\
\bottomrule
\end{tabular}
\end{table*}

\begin{figure*}[htp]
\centering
\includegraphics[width=0.9\textwidth]{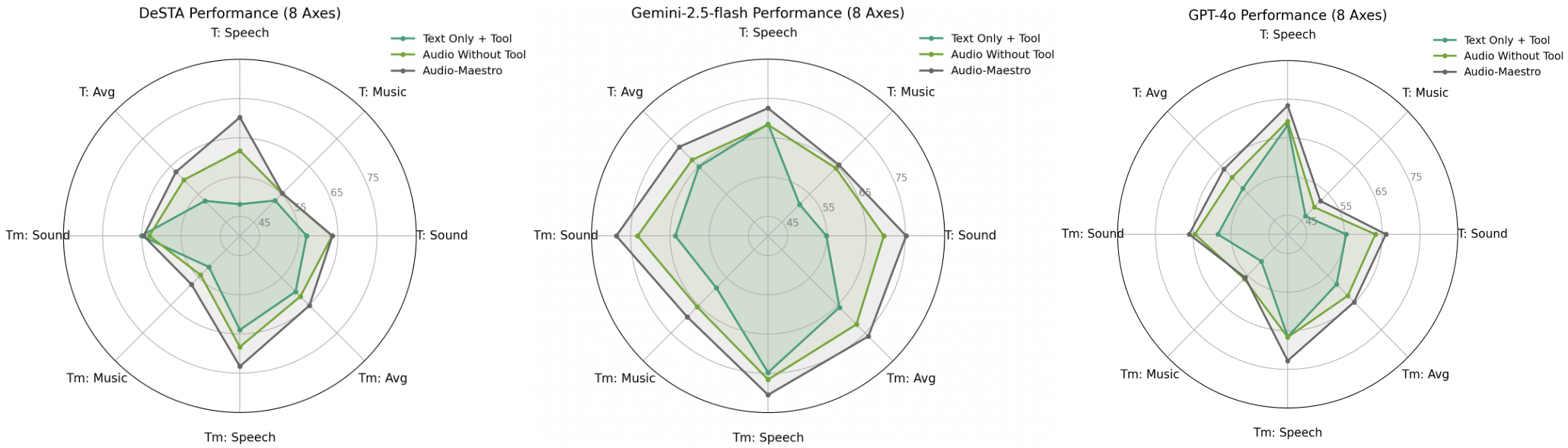}
\caption{\textbf{Performance of Gemini-2.5-flash, DeSTA-2.5, and GPT-4o on the MMAU Benchmark.} The results are segmented into eight categories, and \textbf{T} denotes MMAU-Test and \textbf{Tm} denotes MMAU-Test-Mini.}
\label{fig:result}
\end{figure*}

\section{Result}
\subsection{Main Result}
We evaluate Audio-Maestro under three controlled settings to disentangle the contributions of native audio perception versus tool-augmented reasoning: \textbf{Audio Without Tool} (audio model without external tools), \textbf{Text Only + Tool} (text-only input, equipped with the available tools), and \textbf{Audio-Maestro} (native audio input with the same full tool-calling capabilities). This design explicitly tests whether a natively multimodal agent with direct access to the audio signal outperforms agents that lack native audio perception.

Table~\ref{tab:main-result} shows that Audio-Maestro achieves the best overall accuracy across all models and splits. On \textbf{MMAU-Test}, Audio-Maestro improves the average accuracy compared to the strongest baseline (either Audio Without Tool or Text Only + Tool) across all backbones: from \textbf{67.39\%} to \textbf{71.05\%} on \textbf{Gemini-2.5-flash} (+4.66\%), from \textbf{60.19\%} to \textbf{63.06\%} on \textbf{DeSTA-2.5} (+2.87\%), and from \textbf{60.82\%} to \textbf{63.85\%} on \textbf{GPT-4o} (+3.03\%). Similar gains are observed on \textbf{Test-mini}. These results indicate that empowering the core model with direct access to the raw audio signal (Audio-Maestro) yields superior performance compared to baselines that do not have native audio perception. 

\subsection{Native Perception vs.\ Text-only Tooling}
To quantify the benefit of direct audio perception, we compare \textbf{Text Only + Tool} with \textbf{Audio-Maestro}. Importantly, because the current toolkit does not contain an audio foundation model, the \textbf{Text Only + Tool} agent cannot “listen” to raw audio by calling a built-in audio tool — it must operate from text-only inputs (e.g., prompts, metadata, or transcripts if available) and non-audio utilities. As a result, the text-orchestrated agent is fundamentally unable to access low-level acoustic information.

The results show that this limitation substantially reduces performance: Audio-Maestro outperforms the text+tool agent consistently across backbones (for example, \textbf{Gemini-2.5-flash}: \textbf{58.24\%} $\rightarrow$ \textbf{71.09\%}, a \textbf{7.11\%} absolute gain; \textbf{DeSTA-2.5}: \textbf{52.61\%} $\rightarrow$ \textbf{63.06\%}, a \textbf{10.45\%} absolute gain; \textbf{GPT-4o}: \textbf{56.80\%} $\rightarrow$ \textbf{63.85\%}, a \textbf{7.05\%} absolute gain). This pattern suggests an information bottleneck: without direct access to the raw waveform and its low-level acoustic cues, a text agent cannot recover the rich features required for more challenging audio reasoning tasks. In contrast, Audio-Maestro processes the raw signal directly while simultaneously leveraging the available tools, enabling a more integrated and accurate reasoning process.

\subsection{Tool Invocation Contribution}
We further isolate the impact of tool augmentation by comparing \textbf{Audio Without Tool} against \textbf{Audio-Maestro}, where both settings observe the same raw audio but differ in tool availability. Across all three backbone models, equipping the native audio model with tools yields consistent improvements on MMAU-Test: \textbf{Gemini-2.5-flash} improves from \textbf{67.39\%} to \textbf{71.09\%} (+4.66\%), \textbf{DeSTA-2.5} from \textbf{60.19\%} to \textbf{63.06\%} (+2.87\%), and \textbf{GPT-4o} from \textbf{60.82\%} to \textbf{63.85\%} (+3.03\%). These gains confirm that tools—when combined with native audio perception—effectively offload specialized operations (e.g., precise retrieval, structured analysis), allowing the LALM to focus on higher-level multimodal reasoning. The benefit of tool augmentation therefore complements, rather than substitutes for, native audio perception.

\subsection{Analysis of Error Cases}
\label{sec:error_analysis}
To investigate the bottleneck of Audio-Maestro, we conducted a manual error analysis. For each model, we randomly sample 30 error cases, and provide 3 annotators the original audio and query, tool descriptions, model outputs, the tool’s JSON output, and the ground truth. Each case was categorized into one of three error types: (1) \textbf{Tool Output Error} — incorrect or incomplete tool response; (2) \textbf{Incorrect Tool Selection} — use of an irrelevant or suboptimal tool; (3) \textbf{Result Misinterpretation Error} — misinterpretation of correct tool results. Annotators identified the primary error source and resolved ambiguities through consensus discussion.

We present the result in Table~\ref{tab:error_analysis}, which reveals a clear trend: the majority of failures across all models are attributed to ''Tool Output Errors''. For Gemini, this accounts for 90.0\%. This finding suggests that while the LALMs' ability to select and reason with tools is relatively robust, the primary bottleneck for our framework is the reliability of the external tools themselves. Improving the performance of the underlying specialized models is therefore a critical direction for future work.

\begin{table}[h!]
\centering
\caption{\textbf{Distribution of Error Types}. The analysis is based on 30 randomly sampled error cases for each model. \textbf{TOE} refer to tool output error; \textbf{ICT} means incorrect tool selection; \textbf{RME} refer to result misinterpretation error.}
\label{tab:error_analysis}
\footnotesize
\resizebox{0.8\columnwidth}{!}{%
\begin{tabular}{@{}lccc@{}}
\toprule
\textbf{Model} & \textbf{TOE} & \textbf{ICT} & \textbf{RME} \\
\midrule
DeSTA-2.5    & 73.3\% &  16.7\% & 10.0\% \\
Gemini-2.5-flash   & 90.0\% & 6.7\%  & 3.3\%  \\
GPT-4o   & 80.0\% & 13.3\% & 6.7\% \\
\bottomrule
\end{tabular}%
}
\end{table}

% ---------- Conclusion ----------
\section{Conclusion}
We introduced \textbf{Audio-Maestro}, a tool-augmented framework that enhances LALMs by delegating specialized signal analysis tasks to external tools. Experiments on the MMAU benchmark show that this modular design substantially improves reasoning accuracy across multiple state-of-the-art models. Error analysis further indicates that many failures arise from inaccurate tool outputs, highlighting tool robustness as a key direction for improvement. Our findings emphasize the importance of bridging high-level semantic reasoning with reliable low-level signal operations.

\section{Limitations}
While our tool-augmented audio reasoning framework improves task performance, we acknowledge two main limitations. First, integrating external tools increases inference time, which may limit real-time applications. Second, the framework’s performance depends on tool accuracy. Errors in external tools can propagate to the final output, as confirmed by our manual error case analysis (Table~\ref{tab:error_analysis}). These observations suggest future directions: optimizing tool invocation and enhancing tool robustness.

\section{Generative AI Use Disclosure}
In accordance with ISCA policy, the authors disclose the use of generative AI tools during the preparation of this manuscript. Specifically, these tools were used exclusively for editing, polishing the language, and assisting with LaTeX formatting. No generative AI tools were used to produce any significant conceptual or technical part of the manuscript, nor do they qualify as co-authors. All human authors assume full responsibility and accountability for the final content, results, and scientific integrity of this work, and have consented to its submission.

\bibliographystyle{IEEEtran}
\bibliography{custom}

@article{sakshi2024mmau,
  title={MMAU: A Massive Multi-Task Audio Understanding and Reasoning Benchmark},
  author={Sakshi, S and Tyagi, Utkarsh and Kumar, Sonal and Seth, Ashish and Selvakumar, Ramaneswaran and Nieto, Oriol and Duraiswami, Ramani and Ghosh, Sreyan and Manocha, Dinesh},
  journal={CoRR},
  year={2024}
}

@article{paranjape2023art,
  title={Art: Automatic multi-step reasoning and tool-use for large language models},
  author={Paranjape, Bhargavi and Lundberg, Scott and Singh, Sameer and Hajishirzi, Hannaneh and Zettlemoyer, Luke and Ribeiro, Marco Tulio},
  journal={arXiv preprint arXiv:2303.09014},
  year={2023}
}

@inproceedings{schick2023toolformer,
  title={Toolformer: language models can teach themselves to use tools},
  author={Schick, Timo and Dwivedi-Yu, Jane and Dess{\'\i}, Roberto and Raileanu, Roberta and Lomeli, Maria and Hambro, Eric and Zettlemoyer, Luke and Cancedda, Nicola and Scialom, Thomas},
  booktitle={Proceedings of the 37th International Conference on Neural Information Processing Systems},
  pages={68539--68551},
  year={2023}
}

@inproceedings{kuan2024speech,
  title={Speech-copilot: Leveraging large language models for speech processing via task decomposition, modularization, and program generation},
  author={Kuan, Chun-Yi and Yang, Chih-Kai and Huang, Wei-Ping and Lu, Ke-Han and Lee, Hung-yi},
  booktitle={2024 IEEE Spoken Language Technology Workshop (SLT)},
  pages={1060--1067},
  year={2024},
  organization={IEEE}
}

@inproceedings{chen2024visualcot,
  title     = {Visual Chain-of-Thought Prompting for Knowledge-Based Visual Reasoning},
  author    = {Chen, Z. and Zhou, Q. and Shen, Y. and Hong, Y. and Sun, Z. and Gutfreund, D. and Gan, C.},
  booktitle = {Proceedings of the AAAI Conference on Artificial Intelligence},
  year      = {2024},
  pages     = {1254--1262},
}

@inproceedings{
qin2024toolllm,
title={Tool{LLM}: Facilitating Large Language Models to Master 16000+ Real-world {API}s},
author={Yujia Qin and Shihao Liang and Yining Ye and Kunlun Zhu and Lan Yan and Yaxi Lu and Yankai Lin and Xin Cong and Xiangru Tang and Bill Qian and Sihan Zhao and Lauren Hong and Runchu Tian and Ruobing Xie and Jie Zhou and Mark Gerstein and dahai li and Zhiyuan Liu and Maosong Sun},
booktitle={The Twelfth International Conference on Learning Representations},
year={2024},
url={https://openreview.net/forum?id=dHng2O0Jjr}
}

@article{tseng2025taste,
  title={TASTE: Text-Aligned Speech Tokenization and Embedding for Spoken Language Modeling},
  author={Tseng, Liang-Hsuan and Chen, Yi-Chang and Lee, Kuan-Yi and Shiu, Da-Shan and Lee, Hung-yi},
  journal={arXiv preprint arXiv:2504.07053},
  year={2025}
}

@article{lu2024desta,
  title={DeSTA: Enhancing Speech Language Models through Descriptive Speech-Text Alignment},
  author={Lu, Ke-Han and Chen, Zhehuai and Fu, Szu-Wei and Huang, He and Ginsburg, Boris and Wang, Yu-Chiang Frank and Lee, Hung-yi},
  journal={CoRR},
  year={2024}
}

@article{wu2025step,
  title={Step-audio 2 technical report},
  author={Wu, Boyong and Yan, Chao and Hu, Chen and Yi, Cheng and Feng, Chengli and Tian, Fei and Shen, Feiyu and Yu, Gang and Zhang, Haoyang and Li, Jingbei and others},
  journal={arXiv preprint arXiv:2507.16632},
  year={2025}
}

@article{wu2025audio,
  title={Audio-thinker: Guiding audio language model when and how to think via reinforcement learning},
  author={Wu, Shu and Li, Chenxing and Wang, Wenfu and Zhang, Hao and Wang, Hualei and Yu, Meng and Yu, Dong},
  journal={arXiv preprint arXiv:2508.08039},
  year={2025}
}

@article{deshmukh2025mellow,
  title={Mellow: a small audio language model for reasoning},
  author={Deshmukh, Soham and Dixit, Satvik and Singh, Rita and Raj, Bhiksha},
  journal={arXiv preprint arXiv:2503.08540},
  year={2025}
}

@article{zhong2025omni,
  title={Omni-R1: Reinforcement Learning for Omnimodal Reasoning via Two-System Collaboration},
  author={Zhong, Hao and Zhu, Muzhi and Du, Zongze and Huang, Zheng and Zhao, Canyu and Liu, Mingyu and Wang, Wen and Chen, Hao and Shen, Chunhua},
  journal={arXiv preprint arXiv:2505.20256},
  year={2025}
}

@article{li2025reinforcement,
  title={Reinforcement Learning Outperforms Supervised Fine-Tuning: A Case Study on Audio Question Answering},
  author={Li, Gang and Liu, Jizhong and Dinkel, Heinrich and Niu, Yadong and Zhang, Junbo and Luan, Jian},
  journal={arXiv preprint arXiv:2503.11197},
  year={2025},
  url={https://github.com/xiaomi-research/r1-aqa; https://huggingface.co/mispeech/r1-aqa}
}

@article{hurst2024gpt,
  title={Gpt-4o system card},
  author={Hurst, Aaron and Lerer, Adam and Goucher, Adam P and Perelman, Adam and Ramesh, Aditya and Clark, Aidan and Ostrow, AJ and Welihinda, Akila and Hayes, Alan and Radford, Alec and others},
  journal={arXiv preprint arXiv:2410.21276},
  year={2024}
}

@inproceedings{tangsalmonn,
  title={SALMONN: Towards Generic Hearing Abilities for Large Language Models},
  author={Tang, Changli and Yu, Wenyi and Sun, Guangzhi and Chen, Xianzhao and Tan, Tian and Li, Wei and Lu, Lu and MA, Zejun and Zhang, Chao},
  booktitle={The Twelfth International Conference on Learning Representations},
 year={2023}
}

@article{wang2023neural,
  title={Neural codec language models are zero-shot text to speech synthesizers},
  author={Wang, Chengyi and Chen, Sanyuan and Wu, Yu and Zhang, Ziqiang and Zhou, Long and Liu, Shujie and Chen, Zhuo and Liu, Yanqing and Wang, Huaming and Li, Jinyu and others},
  journal={arXiv preprint arXiv:2301.02111},
  year={2023}
}

@article{xie2025audio,
  title={Audio-reasoner: Improving reasoning capability in large audio language models},
  author={Xie, Zhifei and Lin, Mingbao and Liu, Zihang and Wu, Pengcheng and Yan, Shuicheng and Miao, Chunyan},
  journal={arXiv preprint arXiv:2503.02318},
  year={2025}
}

@article{team2023gemini,
  title={Gemini: a family of highly capable multimodal models},
  author={Team, Gemini and Anil, Rohan and Borgeaud, Sebastian and Alayrac, Jean-Baptiste and Yu, Jiahui and Soricut, Radu and Schalkwyk, Johan and Dai, Andrew M and Hauth, Anja and Millican, Katie and others},
  journal={arXiv preprint arXiv:2312.11805},
  year={2023}
}

@article{ma2023emotion2vec,
  title={emotion2vec: Self-supervised pre-training for speech emotion representation},
  author={Ma, Ziyang and Zheng, Zhisheng and Ye, Jiaxin and Li, Jinchao and Gao, Zhifu and Zhang, Shiliang and Chen, Xie},
  journal={arXiv preprint arXiv:2312.15185},
  year={2023}
}

@inproceedings{mfa2017,
  title     = {Montreal Forced Aligner: Trainable Text-Speech Alignment Using Kaldi},
  author    = {Michael McAuliffe and Michaela Socolof and Sarah Mihuc and Michael Wagner and Morgan Sonderegger},
  year      = {2017},
  booktitle = {Interspeech 2017},
  doi       = {10.21437/Interspeech.2017-1386},
}

@inproceedings{whisper,
  title={Robust speech recognition via large-scale weak supervision},
  author={Radford, Alec and Kim, Jong Wook and Xu, Tao and Brockman, Greg and McLeavey, Christine and Sutskever, Ilya},
  booktitle={International conference on machine learning},
  pages={28492--28518},
  year={2023},
  organization={PMLR}
}

@inproceedings{gong2021ast,
  title = {AST: Audio Spectrogram Transformer},
  author = {Yuan Gong and Yu-An Chung and James Glass},
  booktitle = {Interspeech },
  year = {2021},
  url = {https://arxiv.org/abs/2104.01778}
}

@inproceedings{pyannote,
  author={Hervé Bredin},
  title={{pyannote.audio 2.1 speaker diarization pipeline: principle, benchmark, and recipe}},
  year=2023,
  booktitle={Proc. INTERSPEECH 2023},
}

@inproceedings{McFee2015librosaAA,
  title={librosa: Audio and Music Signal Analysis in Python},
  author={Brian McFee and Colin Raffel and Dawen Liang and Daniel P. W. Ellis and Matt McVicar and Eric Battenberg and Oriol Nieto},
  booktitle={SciPy},
  year={2015},
  url={https://api.semanticscholar.org/CorpusID:33504}
}

@inproceedings{lavechin2023brouhaha,
  title = {Brouhaha: Multi-task training for voice activity detection, speech-to-noise ratio, and C50 room acoustics estimation},
  author = {Marvin Lavechin and Marianne M{\'e}tais and Hadrien Titeux and Alodie Boissonnet and Jade Copet and Morgane Rivi{\`e}re and Elika Bergelson and Alejandrina Cristia and Emmanuel Dupoux and Herv{\'e} Bredin},
  booktitle = {ASRU 2023 (arXiv preprint arXiv:2210.13248)},
  year = {2023},
  url = {https://arxiv.org/abs/2210.13248}
}

@inproceedings{bayron2021autochord,
  author       = {Christopher John Bayron},
  title        = {autochord: Automatic Chord Recognition Library and Chord Visualization App},
  booktitle    = {Extended Abstracts for the Late-Breaking Demo Session of the 22nd International Society for Music Information Retrieval Conference (ISMIR)},
  year         = {2021},
  address      = {Online},
}

\end{document}